\documentclass[floatfix,aps,prd,amsmath,nofootinbib,preprintnumbers,twocolumn]{revtex4}

\usepackage{graphicx}
\usepackage{bm}

\def\half{{1\over 2}}

\def\half{{1\over 2}}
\def\({\left(}
\def\){\right)}
\def\[{\left[}
\def\]{\right]}

\def\e{\begin{equation}}
\def\q{\end{equation}}
\def\m{\begin{eqnarray}}
\def\n{\end{eqnarray}}

\begin{document}

\title{Measuring the spectral running from cosmic microwave background and primordial black holes}

\author{Jun Li$^{1,2}$\footnote{lijun@itp.ac.cn} and Qing-Guo Huang$^{1,2,3,4}$ \footnote{huangqg@itp.ac.cn}}
\affiliation{$^1$ CAS Key Laboratory of Theoretical Physics,\\ Institute of Theoretical Physics, \\Chinese Academy of Sciences, Beijing 100190, China\\
$^2$ School of Physical Sciences, \\University of Chinese Academy of Sciences,\\ No. 19A Yuquan Road, Beijing 100049, China\\
$^3$ Center for Gravitation and Cosmology, College of Physical Science and Technology, Yangzhou University, Yangzhou 225009, China\\
$^4$ Synergetic Innovation Center for Quantum Effects and Applications, Hunan Normal University, Changsha 410081, China}

\date{\today}

\begin{abstract}

We constrain the spectral running by combining cosmic microwave background (CMB) data, baryon acoustic oscillation (BAO) data and the constraint from primordial black holes (PBHs). We find that the constraint from PBHs has a significant impact on the running of running of scalar spectral index, and a power-law scalar power spectrum without running is consistent with observational data once the constraint from PBHs is taken into account. In addition, from the constraints on the slow-roll parameters, the derived tensor spectral index in the single-field slow-roll inflation model is quite small, namely $|n_t|\lesssim 9.3\times 10^{-3}$ which will be very difficult to be measured by CMB data only in the future, and the absolute value of derived running of tensor spectral index is not larger than $2.1\times 10^{-4}$ at $95\%$ confidence level.


\end{abstract}

\pacs{???}

\maketitle


\section{Introduction}

The current cosmological observations, for example the Planck satellite \cite{Ade:2015xua,Ade:2015lrj}, provide strong evidence for the standard six-parameter $\Lambda$CDM model. However, when the base model is extended and other cosmological parameters are allowed to freely vary, a few anomalies are present. For example, the running of running of the scalar spectral index which is defined by
\m
\beta_s\equiv {d^2 n_s\over d \ln k^2}
\n
reads
\m
\beta_s=0.025\pm 0.013
\label{bs1}
\n
from Planck data in \cite{Ade:2015lrj} and
\m
\beta_s=0.021\pm0.013
\label{bs2}
\n
from the combination of Planck \cite{Ade:2015lrj}, BK14 \cite{Array:2015xqh} and Baryon Acoustic Oscillation (BAO) \cite{Gil-Marin:2015nqa,Beutler:2011hx,Ross:2014qpa} data in \cite{Huang:2015cke} at $68\%$ confidence level (CL) on the cosmic scales, where $n_s$ is the spectral index of scalar power spectrum.

In 1971, Hawking proposed that the highly overdense region of inhomogeneities would eventually cease expanding and collapse into a black hole in \cite{Hawking:1971ei}. A large positive value of the running of running may lead to the formation of primordial black holes (PBH) at small scales in the early universe.  Different from the astrophysical black holes which should be heavier than a particular mass (around 3 solar mass $M_\odot$), PBHs can have very small masses. However, the PBHs with masses smaller than $10^{-18} M_\odot$ would have completely evaporated by now due to the Hawking radiation. In particular, the quantum emissions from PBHs with mass around $10^{-18} M_\odot$ can generate a $\gamma$-ray background which should be observed today. Conversely, the non-detection of such a $\gamma$-ray background in our Universe provides a stringent constraint on the PBHs with such a mass. Recently various constraints on the abundance of PBHs are present in \cite{Chen:2016pud,Green:2016xgy,Schutz:2016khr,Wang:2016ana,Gaggero:2016dpq,Ali-Haimoud:2016mbv,Blum:2016cjs,Horowitz:2016lib,Kuhnel:2017pwq,Inoue:2017csr,Carr:2017jsz,Green:2017qoa,Zumalacarregui:2017qqd,Chen:2018czv}. Ones expect that the absence of PBHs should tightly constrain the running of running of scalar spectral index.


On the other hand, inflation \cite{Guth:1980zm,Starobinsky:1980te,Linde:1981mu,Albrecht:1982wi} is taken as the leading paradigm for the physics in the very early universe. The initial inhomogenieties and spatial curvature are supposed to be stretched away by the quasi-exponential expansion of inflation. Because the Hubble parameter is roughly a constant during inflation, it predicts a nearly scale-invariant power spectrum of the curvature (scalar) perturbations seeded by the quantum fluctuations of inflaton field during inflation. Typically, the scalar spectral index $n_s$ is related to the number of e-folds $N$ before the end of inflation by $n_s-1\sim -1/N$, and then $\beta_s\sim -1/N^3\sim (n_s-1)^3\sim -{\cal O}(10^{-5})$. In this sense, the running of running of scalar spectra index given in Eqs.~(\ref{bs1}) and (\ref{bs2}) seems too large to fit in the typical inflation models.


In this paper we adopt two methods to compare inflationary predictions with current cosmological datasets, in particular including the constraint from the PBHs. The first method consists of a phenomenological parameterization of the primordial spectra of both scalar and tensor perturbations, and the second exploits the analytic slow-roll-parameter dependence of primordial perturbations.

This paper is organized as follows. In Sec.~2, we explain the datasets used in this paper. In Sec.~3 and Sec.~4, we present the constraints on the phenomenological parameters and the slow-roll parameters with the publicly available codes CosmoMC \cite{Lewis:2002ah}, respectively. A brief summary will be given in Sec.~5.

\section{Data}

The full-mission Planck observes the temperature and polarization anisotropies of the cosmic microwave background (CMB) radiation. In this paper we also combine all the data taken by the BICEP2 and Keck Array CMB polarization experiments up to and including the 2014 observing season \cite{Array:2015xqh} with the Planck data.

Baryon Acoustic Oscillation (BAO) detections measure the correlation function and power spectrum in the clustering of galaxies. Measuring the position of these oscillations in the matter power spectra at different redshifts removes degeneracies in the interpretation of the CMB anisotropies.
The BAO data adopted in this paper include 6dFGS \cite{Beutler:2011hx}, MGS \cite{Ross:2014qpa}, BOSS $\mathrm{DR11}\_{\mathrm{Ly}\alpha}$ \cite{Delubac:2014aqe}, BOSS DR12 with nine anisotropic measurements \cite{Wang:2016wjr}, and eBOSS DR14 \cite{Ata:2017dya}.

The mass of PBH is roughly given by the horizon mass at the time of formation, namely
\m
m=\gamma {4\pi\over 3}\rho H^{-3},
\n
where $\gamma=3^{-3/2}\simeq 0.2$ \cite{Carr:1975qj}. In the comoving units, the horizon scale is $R=(aH)^{-1}\propto a$ during radiation domination. Since constant-entropy expansion implies $T\propto g_*^{-1/3} a^{-1}$, $\rho\propto g_*T^4\propto g_*^{-1/3} a^{-4}$ and then the mass of PBH is $m\propto g_*^{-1/3} a^2$, where the number of relativistic degrees of freedom $g_*$ is around 3 at equality and $10^2$ in the early universe. Therefore
\m
{m\over M_\odot}\simeq (k_\odot R)^2,
\n
where $k_\odot\simeq 10^6$ Mpc$^{-1}$ is the perturbation mode which re-enter the horizon when the one-solar mass PBH can be formed. Therefore the scale correspond to the formation of PBH with a mass $10^{-18}M_\odot$ is roughly given by $k_c\simeq 10^{15}$ Mpc$^{-1}$. Actually the formation process of the PBHs is still poorly understood. As a conservative estimation, we require that the scalar power spectrum should smaller than the unity at $k=k_c$, namely
\m
P_s(k_c)\leq 1.
\label{pspbh}
\n
Otherwise, a lot of PBHs with mass $10^{-18}M_\odot$ should be formed in the early Universe and generate an observable $\gamma$-ray background. We take Eq.~(\ref{pspbh}) as the constraint on the scalar power spectrum from PBHs.

\section{Constraints on the spectral running from CMB and PBHs}

In this section, the power spectra of the scalar and tensor perturbations are parameterized by
\m
P_s(k)&=&A_s\(\frac{k}{k_*}\)^{n_s-1+\frac{1}{2}\alpha_s\ln(k/k_*)+\frac{1}{6}\beta_s(\ln(k/k_*))^2+...}, \label{eqs:spectrumscalar}\\
P_t(k)&=&A_t\(\frac{k}{k_*}\)^{n_t+\frac{1}{2}\alpha_t\ln(k/k_*)+...},\label{eqs:spectrumtensor}
\n
where $A_s(A_t)$ is the scalar (tensor) amplitude at the pivot scale $k_*=0.05$ Mpc$^{-1}$, $n_s$ is the scalar spectral index, $\alpha_s\equiv\mathrm{d} n_s/\mathrm{d}\ln k$ is the running of scalar spectral index, $\beta_s\equiv{\mathrm{d}^2n_s}/{\mathrm{d}\ln k^2}$ is the running of running of scalar spectral index, $n_t$ is the tensor spectral index, and $\alpha_t\equiv \mathrm{d} n_t/\mathrm{d}\ln k$ is the running of tensor spectral index. Usually we introduce a new parameter, namely the tensor-to-scalar ratio $r$, to quantify the tensor amplitude compared to the scalar amplitude at the pivot scale:
\e
r\equiv\frac{A_t}{A_s}.
\q
It is known that the single parameter Harrison-Zeldovich spectrum $(n_s=1)$ does not fit the data and at least the parameters $A_s$ and $n_s$ in the expansion of the primordial scalar power spectrum are needed. Here the spectral index of tensor power spectrum is set as $n_t=-r/8$ which is nothing but the consistency relation to lowest order in the single-field slow-roll inflation model, and $\alpha_t=0$.

\begin{figure}
\centering
\includegraphics[width=8.8cm]{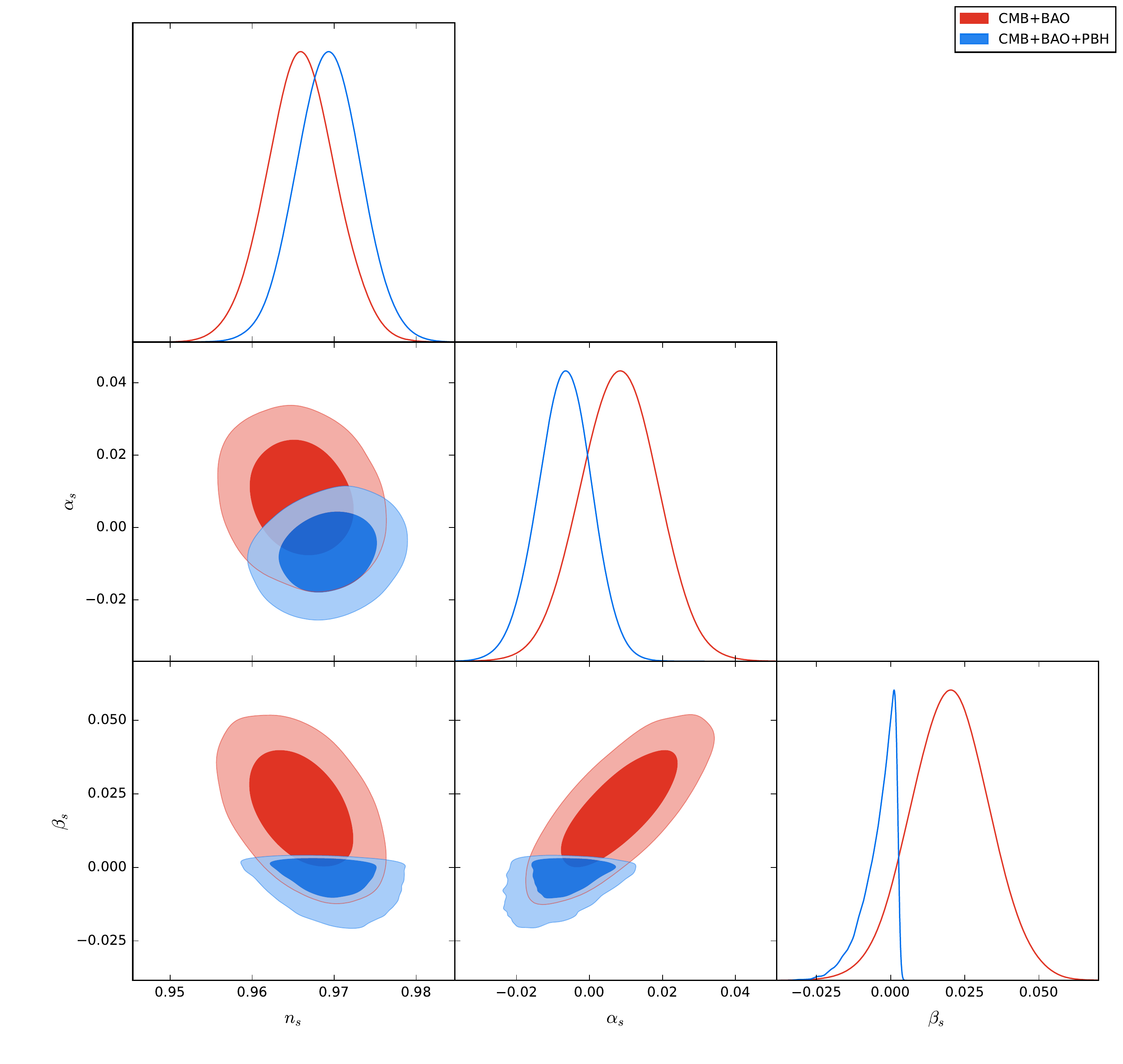}
\caption{The contour plots and the likelihood distributions for $n_s$, $\alpha_s$, $\beta_s$ at the $68\%$ and $95\%$ CL by using CMB+BAO and CMB+BAO+PBH, respectively.}
\label{fig:spectrum}
\end{figure}

We consider the six parameters in the standard $\Lambda$CDM model, i.e. the baryon density parameter $\Omega_b h^2$, the cold dark matter density $\Omega_c h^2$, the angular size of the horizon at the last scattering surface $\theta_\text{MC}$, the optical depth $\tau$, the scalar amplitude $A_s$ and the scalar spectral index $n_s$. We extend this scenario by adding the running of the scalar spectral index $\alpha_s$, the running of running $\beta_s$ and the tensor amplitude, or equivalently the tensor-to-scalar ratio $r$. We constrain all of these 9 parameters in the $\Lambda$CDM+$r+\alpha_s+\beta_s$ model by adopting two different data combinations, namely CMB+BAO and CMB+BAO+PBH, respectively. The results are given in Table.~\ref{table:spectra} and Fig.~\ref{fig:spectrum}.
\begin{table}
\newcommand{\tabincell}[2]{\begin{tabular}{@{}#1@{}}#2\end{tabular}}
  \centering
  \begin{tabular}{c | c | c}
  \hline
  \hline
  Parameter & \tabincell{c}{CMB+BAO}& \tabincell{c}{CMB+BAO+PBH} \\
  \hline
  $\Omega_bh^2$ & $0.02232\pm0.00015$ & $0.02238\pm0.00014$ \\
  $\Omega_ch^2$ & $0.1180\pm0.0008$ & $0.1179\pm0.0008$ \\
  $100\theta_{\mathrm{MC}}$ & $1.0410\pm0.0003$ & $1.0410\pm0.0003$ \\
  $\tau$ & $0.079\pm0.012$ & $0.073\pm0.012$ \\
  $\ln\(10^{10}A_s\)$ & $3.088\pm0.023$ & $3.077\pm0.022$ \\
  $n_s$ & $0.9660\pm0.0040$ & $0.9693\pm0.0038$ \\
  $\alpha_s$ & $0.0083^{+0.0104}_{-0.0103}$ & $-0.0067^{+0.0073}_{-0.0074}$ \\
  $\beta_s$ & $0.020\pm 0.013$ & $-0.0038^{+0.0069}_{-0.0025}$ \\
  $r$ ($95\%$ C.L.) & $<0.087 $ & $<0.079$  \\
  \hline
  \hline
  \end{tabular}
  \caption{The $68\%$ limits on the cosmological parameters in the $\Lambda$CDM+$r+\alpha_s+\beta_s$ model from the data combinations of CMB+BAO and CMB+BAO+PBH respectively. }
  \label{table:spectra}
\end{table}
The constraint on the running of running of scalar spectral index is
\m
\beta_s=0.020\pm 0.013
\n
at $68\%$ CL from CMB+BAO, and
\m
\beta_s=-0.0038^{+0.0069}_{-0.0025}
\n
at $68\%$ CL from CMB+BAO+PBH. We see that the constraint on the running of running of scalar spectral index is significantly affected by adding the constraint of PBHs, and a power-law scalar power spectrum without running is consistent with the data once the constraint from PBHs is taken into account.


\section{Constraints on the slow-roll parameters}

In this section, we focus on the cannonical single-field slow-roll inflation model in which the inflation is driven by the inflaton potential $V(\phi)$. The dynamics of inflation is govern by
\m
H^2={1\over 3 M_p^2}\[\half {\dot\phi}^2+V(\phi)\], \\
\ddot \phi+3H\dot\phi+V^\prime (\phi)=0,
\n
where $M_p=1/\sqrt{8\pi G}$ is the reduced Planck energy scale and the dot and prime denote the derivative with respective to the cosmic time $t$ and the inflaton field $\phi$, respectively. The inflaton field slowly rolls down its potential if $\epsilon\ll 1$ and $|\eta|\ll 1$, where
\m
\epsilon&=&\frac{M_p^2}{2}\(\frac{V^\prime\(\phi\)}{V\(\phi\)}\)^2,\\
\eta&=&M_p^2\frac{V^{\prime\prime}\(\phi\)}{V\(\phi\)}.
\n
The amplitude of scalar and tensor perturbation power spectra are given by
\m
P_s&\simeq& \[1+{25-9c\over 6}\epsilon-{13-3c\over 6}\eta\] {V/M_p^4\over 24\pi^2 \epsilon}, \\
P_t&\simeq& \[1-{1+3c\over 6}\epsilon\]{V/M_p^4\over 3\pi^2/2},
\n
in \cite{Huang:2014yaa}, where $c\simeq 0.08145$. See \cite{Stewart:1993bc,Huang:2006yt} as well. And hence, we have
\m
r&\approx&16\epsilon\[1-\frac{13-3c}{6}\(2\epsilon-\eta\)\],\label{slr}\\
n_t&\approx&-2\epsilon-\frac{2\(2+3c\)}{3}\epsilon^2-\frac{1-3c}{3}\epsilon\eta,\label{slnt}\\
\nonumber\alpha_t&\approx&-8\epsilon^2+4\epsilon\eta-\frac{8\(5+6c\)}{3}\epsilon^3+2\(1+7c\)\epsilon^2\eta\\&&+2\(1-c\)\epsilon\eta^2,\label{slalphat}\\
\nonumber n_s&\approx&1-6\epsilon+2\eta+\frac{2\(22-9c\)}{3}\epsilon^2-2\(7-2c\)\epsilon\eta\\&&+\frac{2}{3}\eta^2,\\
\nonumber\alpha_s&\approx&-24\epsilon^2+16\epsilon\eta-2\xi+\frac{8\(41-18c\)}{3}\epsilon^3\\&&\nonumber-\frac{4\(109-36c\)}{3}\epsilon^2\eta+4\(9-2c\)\epsilon\eta^2
\\&&+2\(11-3c\)\epsilon\xi-\frac{25-3c}{6}\eta\xi,\\
\nonumber\beta_s&\approx&-192\epsilon^3+192\epsilon^2\eta-32\epsilon\eta^2-24\epsilon\xi+2\eta\xi+2\sigma\\&&\nonumber+96\(13-6c\)\epsilon^4
-\frac{8\(791-288c\)}{3}\epsilon^3\eta\\&&\nonumber+\frac{16\(173-48c\)}{3}\epsilon^2\eta^2-\frac{8\(31-6c\)}{3}\epsilon\eta^3\\&&\nonumber
+\frac{4\(235-72c\)}{3}\epsilon^2\xi-\frac{511-111c}{3}\epsilon\eta\xi\\&&\nonumber
+\frac{29-3c}{6}\eta^2\xi+\frac{25-3c}{6}\xi^2-\frac{103-27c}{3}\epsilon\sigma\\
&&+\frac{55-9c}{6}\eta\sigma,
\n
where
\m
\xi&=&M_p^4\frac{V^\prime\(\phi\)V^{\prime\prime\prime}\(\phi\)}{V^2\(\phi\)},\\
\sigma&=&M_p^6\frac{{V^\prime}^2\(\phi\)V^{\prime\prime\prime\prime}\(\phi\)}{V^3\(\phi\)}.
\n

\begin{figure}
\centering
\includegraphics[width=8.8cm]{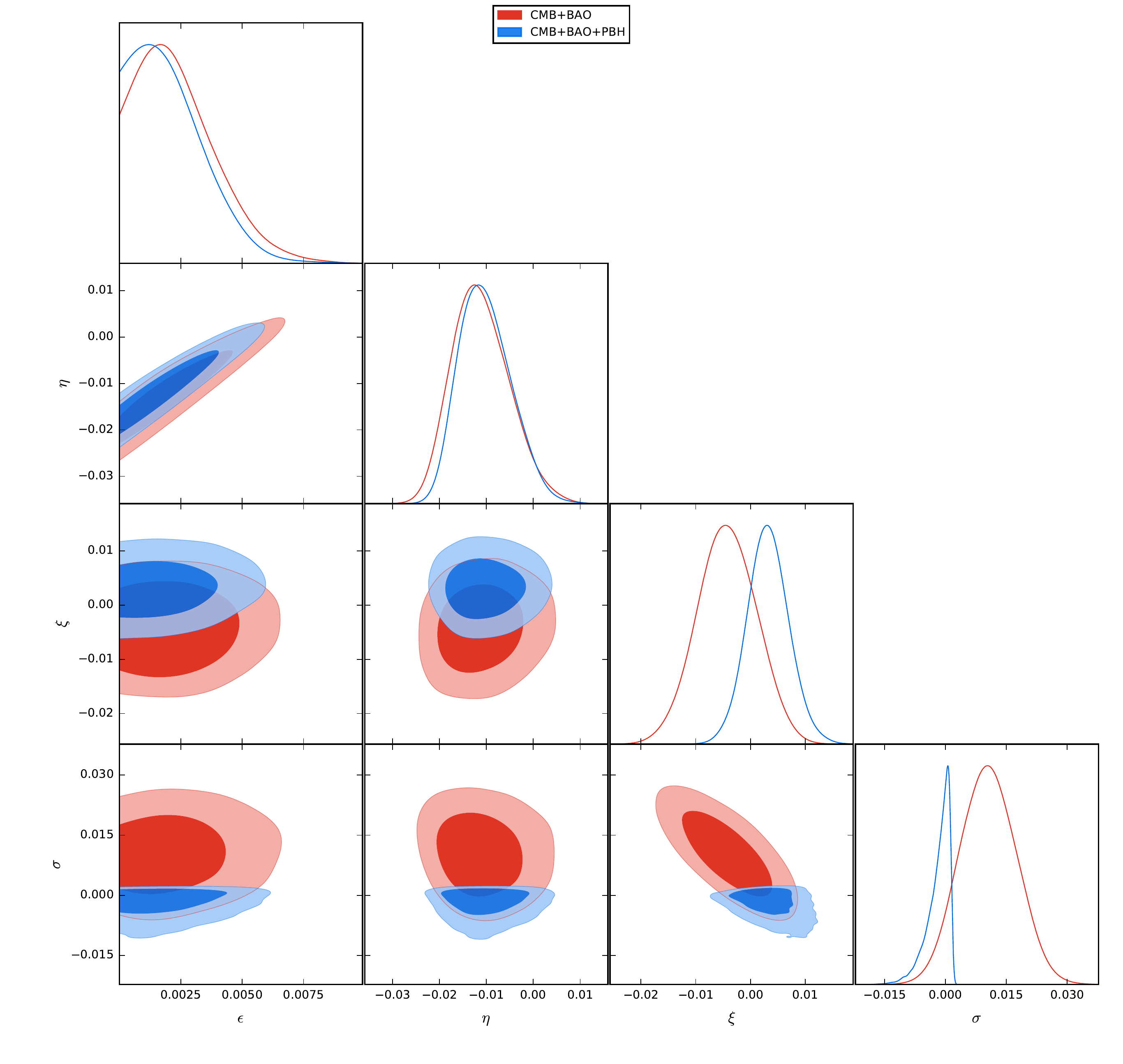}
\caption{The contour plots and the likelihood distributions for the slow-roll parameters $\epsilon$, $\eta$, $\xi$ and $\sigma$ at the $68\%$ and $95\%$ CL from CMB+BAO and CMB+BAO+PBH, respectively.}
\label{fig:slowroll}
\end{figure}

Here the nine parameters sampled in the CosmoMC are $\{\Omega_b h^2, \Omega_c h^2, \tau, \theta_\text{MC}, A_s, \epsilon, \eta, \xi, \sigma\}$, and then $n_s$, $\alpha_s$, $\beta_s$, $r$, $n_t$ and $\alpha_t$ are all taken as the derived parameters. The constraints on the slow-roll parameters $\{\epsilon, \eta, \xi, \sigma\}$ and the contour plots of these slow-roll parameters are illustrated in Table.~\ref{table:slowroll} and Fig.~\ref{fig:slowroll}. We see that the constraint on the fourth slow-roll parameter $\sigma$ is improved once the constraint from PBHs is included.
\setlength{\belowcaptionskip}{1.cm}

\begin{table}

\newcommand{\tabincell}[2]{\begin{tabular}{@{}#1@{}}#2\end{tabular}}
  \centering
  \begin{tabular}{c | c | c}
  \hline
  \hline
  Parameter & \tabincell{c}{CMB+BAO}& \tabincell{c}{CMB+BAO+PBH} \\
  \hline
  $\epsilon$ & $<0.0053$ & $<0.0047$ \\
  $\eta$ & $-0.0112^{+0.0122}_{-0.0116}$ & $-0.0104^{+0.0111}_{-0.0102}$ \\
  $\xi$ & $-0.0043^{+0.0106}_{-0.0107}$ & $0.0031^{+0.0077}_{-0.0074}$ \\
  $\sigma$ & $0.0104^{+0.0134}_{-0.0133}$ & $-0.0018^{+0.0045}_{-0.0062}$ \\
  \hline
  $r$ & $<0.082$ & $<0.073$ \\
  $-n_t\ (\times 10^{-2})$ & $<1.1$ & $<0.93$ \\
  $-\alpha_t\ (\times 10^{-4})$ & $<2.6$ & $<2.1$ \\
  \hline
  \hline
  \end{tabular}
  \caption{The $95\%$ limits on the slow-roll parameters and the derived parameters from the data combinations of CMB+BAO and CMB+BAO+PBH respectively. }
  \label{table:slowroll}
\end{table}

\begin{figure}
\centering
\includegraphics[width=8.8cm]{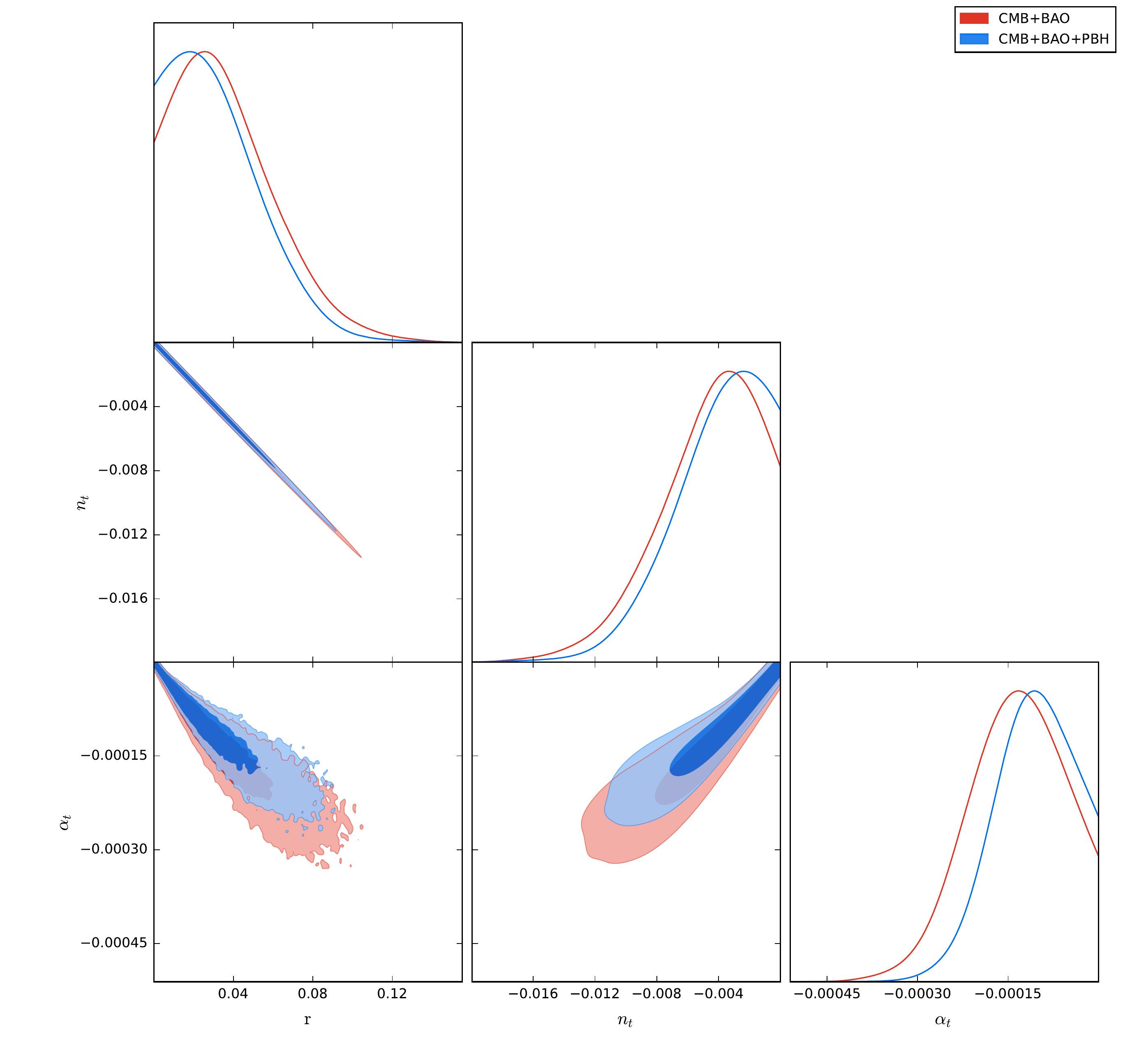}
\caption{The derived parameters $r$, $n_t$ and $\alpha_t$ for the constrained canonical single-field slow-roll inflation model at the $68\%$ and $95\%$ CL from CMB+BAO or CMB+BAO+PBH.}
\label{fig:slowroll_fig}
\end{figure}

Here we also work out the predictions of the slow-roll inflation model constrained by the observational data. The parameters $\{r, n_t, \alpha_t\}$ characterizing the tensor power spectrum can be obtained by adopting the Eqs. ~(\ref{slr}), (\ref{slnt}) and (\ref{slalphat}). The results are showed in Table.~\ref{table:slowroll} and Fig.~\ref{fig:slowroll_fig}.
In particular, we notice that both the derived tensor spectral index $n_t$ and its running $\alpha_t$ are negative in the constrained single-field inflation model, and the constraint on them at $95\%$ C.L. are
\m
-n_t&<& 1.1\times 10^{-2}, \\
-\alpha_t&<& 2.6\times 10^{-4},
\n
from CMB+BAO, and
\m
-n_t&<& 9.3\times 10^{-3}, \\
-\alpha_t&<& 2.1\times 10^{-4},
\n
from CMB+BAO+PBH. However, in \cite{Huang:2017gmp}, the optimistic estimation indicates that the uncertainty at $68\%$ C.L. on the tensor spectral index is $\sigma_{n_t}\simeq 1.1\times 10^{-2}$ due to the cosmic variance only for CMB multipoles less than 300. Because the small-scale CMB B-modes are dominated by CMB lensing, it implies that it is very difficult to measure the tensor spectral index for the single-field slow-roll inflation model by only using the CMB data.




\section{Summary}

In this paper we use two methods to constrain the slow-roll inflation models by combining CMB, BAO and the constraint from PBHs. Even though a positive running of running of scalar spectral index is slightly preferred by the data combination of CMB and BAO datasets, a power-law scalar power spectrum without running is consistent with the data once the constraint from PBHs is taken into account.

We can also directly constrain the slow-roll parameters from the observational data. An advantage of this method is that we can work out the predictions of single-field slow-roll inflation model by using these constrained slow-roll parameters. For example, we illustrate the predictions of the parameters characterizing the tensor power spectrum, and find that both the tensor spectral index and its running are negative and their absolute values are not larger than $9.3\times 10^{-3}$ and $2.1\times 10^{-4}$ at $95\%$ C.L., respectively. Our results imply that it is very difficult to measure these two parameters in the future.

\vspace{5mm}
\noindent {\bf Acknowledgments}.

We acknowledge the use of HPC Cluster of ITP-CAS. This work is supported by grants from NSFC (grant NO. 11335012, 11575271, 11690021, 11747601), Top-Notch Young Talents Program of China, and partly supported by Key Research Program of Frontier Sciences, CAS.



\end{document}